\documentclass[a4paper,12pt]{article}
\usepackage{a4wide}
\usepackage{authblk}
\usepackage{blindtext}
\usepackage{graphicx}
\usepackage{amsmath}
\usepackage{amsfonts}
\usepackage{amssymb}

\title {Connecting the Unstable Region of the Entropy to the Pattern of the Fisher's Zeros Map}

    \author[1]{J. C. S. Rocha}
    \affil[1]{Departamento de F\'isica, ICEB, Universidade Federal de Ouro Preto, CEP 35402-136, Ouro Preto, Minas Gerais, Brazil}

    \author[2]{B.V. Costa}
    \affil[2]{Retired Professor, Laborat\'orio de Simula\c c\~ao, Departamento de F\'isica, ICEx \\ Universidade Federal de Minas Gerais, 31720-901 Belo Horizonte, Minas Gerais, Brazil}

\date{\today}

\begin{document}
\maketitle
\abstract{
Phase transitions are one of the most interesting natural phenomena.
For finite systems, one of the concerns in the topic is how to classify a specific transition as being of first, second, or even of a higher order, according to the Ehrenfest classification.
The partition function provides all the thermodynamic information about the physical systems, and a phase transition can be identified by the complex temperature where it is equal to zero.
In addition, the pattern of the zeros on the complex temperature plane can provide evidence of the order of the transition.
In this manuscript, we present an analytical and simulational study connecting the microcanonical analysis of the unstable region of the entropy to the canonical partition function zeros. We show that, for the first-order transition, the zeros accumulate uniformly in a vertical line on the complex inverse temperature plane as discussed in previous works. We illustrate our calculation using the $147$ particles Lennard-Jones cluster.}
\section{Introduction}
\label{Introduction}
The transitions between different states of matter observed in macroscopic systems, such as the solidification of water into ice or the demagnetization of a magnetic rod, are well-described phenomena. In the case of water freezing, this transition involves the coexistence of both the liquid and solid phases, whereas the demagnetization process does not exhibit a distinct boundary between the ferromagnetic and paramagnetic phases. According to P.~Ehrenfest~\cite{Sauer}, these phase transitions are classified as being of first and second order, respectively. This classification is based on appointing the lowest discontinuous derivative of the free energy at the transition point, thereby implying a non-analytical point in this function. 

In the realm of equilibrium statistical mechanics, the key entity for investigating a system is its partition function. The canonical partition function, denoted as $Z(\beta,V,N)$, holds a connection to the Helmholtz free energy through the limit $F(\beta,v) = -\beta^{-1} \lim_{N\to \infty}\ln{Z(\beta,V,N)} $, while $v = V/N$ tends to a constant. Here, $\beta = 1/k_BT$ represents the canonical inverse temperature, $k_B$ signifies the Boltzmann constant, and $v$ denotes the specific volume. Consequently, the non-analytic behavior of $F(\beta,v)$ occurs at points where $\lim_{N\to \infty} Z(\beta,V,N) = 0$.

The groundbreaking work of Lee~and~Yang~\cite{Lee_Yang} and its extension by Fisher~\cite{Fisher} established that studying the zeros of the partition function offers a rigorous framework for understanding phase transitions~\cite{Ruelle,Wei-Liu,Peng}. Although the partition function is composed exclusively of positive terms, implying the absence of real positive roots for any finite system, inspecting the zeros of small systems allows for the revelation of some properties of the thermodynamic system, for instance, the transition temperature.

Consider the analytical continuation of the partition function $Z=Z(\mathcal{B})$ with $\mathcal{B} = \beta + i\tau$, where $\beta$ represents the inverse temperature and $\tau$ is an imaginary parameter. In the thermodynamic limit, a phase transition occurs at $\lim_{N \rightarrow \infty} Z(\mathcal{B}_k) = 0$ if $\tau_k = 0$. The manner in which the zeros approach this limiting point is indicative of the transition's order.

In the late 1960s, S.~Grossmann and W.~Rosenhauer~\cite{Grossmann1967,Grossmann1969} showed that the phenomenologically known types of phase transition can be characterized by the way that the density of zeros, which is the thermodynamic limit of the distribution of zeros (DOZ), behave toward the transition point. They proposed a general Finite-Size Scale (FSS) method for the DOZ which accumulates in lines that tend to cut the real axis under a certain slope, $\gamma = (\beta - \beta_c)/\tau$, whereas the density function can be described by a simple power law $\phi(\tau) \approx \tau^{\alpha}$. After that, S.~Grossmann and V.~Lehmann~\cite{Lehmann1969} provided some results of this method for realistic physical models.

Towards the end of the twentieth century, P.~Borrmann et al~\cite{Borrmann} proposed a classification scheme for phase transitions in finite systems based on the method introduced by S.~Grossmann and W.~Rosenhauer. Similarly, by analyzing the DOZ they classified the type of the transition by both: the angle of the zeros lines toward the real axis and the distance between the zeros in this line. For a pseudo-first-order phase transition, this line is perpendicular to the real axis and, concomitantly, the zeros are evenly spaced, see Fig.~\ref{LJzerosBetaMap}.

More recently, M.P.Taylor et al\cite{Taylor} empirically demonstrated the connection between the curvature properties of entropy, denoted as $S$, and the DOZ. In the microcanonical analysis, a convex behavior of $S$, i.e. an unstable region, is related to a first-order transition~\cite{QiBachmann}. The double-touching tangent line construction on this convex intruder can define both the energy range of the non-stable region and the transition temperature. The authors calculated the zeros of $Z$ by considering this truncated energy range and  $x=e^{-\mathcal{B} E}$ as a variable and showed that it leads to a circle on the complex $x$ plane map. Solving it for $\mathcal{B}$, this circle leads to a vertical line on the complex $\mathcal{B}$ plane map, which corroborates with P.~Bormann and collaborators' results. Additionally, they observed another pattern of zeros that pinch the real axis, which they attributed to a higher-order transition.

In the present work, we propose an alternative analytical argument to establish a connection between the unstable region of the entropy and the vertical line pattern observed in the Distribution of Zeros (DOZ), as empirically shown by  M.P.~Taylor and collaborators. This paper is organized as follows: in section~\ref{microcanonical}, we present the microcanonical analysis of phase transition. Subsequently, in section~\ref{fishersZeros} we introduce the Fisher zeros and in the section~\ref{Borrmann_section} the classification scheme proposed by P.~Borrmann et al. In section \ref{firstOrderzeros} we outline the analytical arguments that a first-order transition leads to a vertical line pattern of the zeros on the complex $\mathcal{B}$ plane map. Our results are then compared with a Monte Carlo simulation of the $147$ particles Lennard-Jones cluster in section~\ref{LJclusterZeros}. Finally, in section \ref{Final_Remarks} we present our conclusions.

 \section{Methodology}

 \subsection{Microcanonical Ensemble}
\label{microcanonical}
    In the microcanonical approach to statistical mechanics, entropy carries all the information necessary to describe the system. The first probabilistic statement for entropy was made for the ideal gas in 1872 by L.~Boltzmann~\cite{Brush}. In 1901, M.~Planck stated his famous formula,
    \begin{equation}
    S(E) = k_B\ln{\Omega(E)},
    \label{PlanckBoltzmann}
    \end{equation}
    as the expression for the entropy of black bodies~\cite{Planck}, with $\Omega(E)$ standing for the number of ways in which the system can be can be realized with energy $E$. For simplicity, in this work we measure $S$ in units of $k_B$.
    
    In microcanonical statistics, the equilibrium state of a thermodynamic system is defined by derivatives of $S$. For instance, the inverse microcanonical temperature is given by the derivative of entropy $S$ with respect to energy $E$, while keeping the set of independent extensive quantities $\{X\}$ (such as volume $V$, number of particles $N$, magnetization $M$, etc.) constant, i.e.
    \begin{equation}
     \bar{\beta}(E) = \bar{T}^{-1} = \left( \frac{\partial S}{ \partial E}\right)_{\{X\}} .
    \end{equation}
 The overbar is used to emphasize that the quantity is a microcanonical parameter. It is important to note that $\bar{\beta}/k_B$ recovers the usual canonical $\beta$ in the thermodynamic limit.

Considering an energy region without any transition, the function $S(E)$ is strictly monotonically increasing, concave, and positive. Consequently, $\bar{\beta}$ is a monotonically decreasing, convex, and positive function. Higher-order derivatives of entropy, denoted by 
    \begin{eqnarray}
    \bar{\gamma}(E) = \left(\frac{\partial^2 S}{\partial E^2}\right)_{\{X\}}  &
    \text{ \ \ and  \ \  }   &
    \bar{\delta}(E) = \left( \frac{\partial^3 S}{\partial E^3}\right)_{\{X\}},
    \end{eqnarray}
 are increasing concave negative and positive decreasing convex positive functions, respectively, and so on.
    
      A convex behavior of the entropy indicates a non-stable region, so that, a change in the concavity of $S(E)$ corresponds to a first-order phase transition. The touching points of the double-tangent line across the convex region define the latent heat and the energy range of the transition, $[E', E'']$. Additionally, the slope of this line defines the transition temperature, see Fig.~\ref{LJentropy}. 
      
\subsection{Fisher's Zeros}
\label{fishersZeros}
    From the point of view of the Canonical Ensemble, the partition function can be seen as the Laplace transform of $\Omega(E)$. For a system with a continuous energy domain, it can be written as:
    \begin{equation}
    Z(\mathcal{B},V,N) = \int \mathrm{d}E\ \Omega(E)e^{-\mathcal{B} E},
    \label{Z_laplace}
    \end{equation}
    where, $\mathcal{B} =  \beta + i  \tau$ represents the complex inverse temperature and $E$ depends on $V$ and $N$. One can introduce a discretization approach by considering an energy bin of size $\varepsilon$. Consequently, the energy of the $k$-th bin can be expressed as:
    \begin{equation}
    E_k = E_0 + k\varepsilon,
    \label{E_level}
    \end{equation}
    where $E_0$ stands for the ground state energy. In this approach, $\Omega(E_k)$ represents the number of states with energy between $E_k$ and $E_k+\varepsilon$. By considering a discrete version of eq.~(\ref{Z_laplace}) and incorporating the energy given by eq.~(\ref{E_level}), we can express the partition function as:
    \begin{equation}
    Z(\mathcal{B},V,N)  =  e^{-\mathcal{B} E_0} \sum_{k=0}^{n} \Omega_{k} e^{-\mathcal{B} k \varepsilon},
        \label{Z_total}
    \end{equation}
    \noindent
    where $\Omega_k \equiv \Omega(E_k)$ and $n$ is the number of energy bins. Following Fisher we define a new variable
    \begin{equation}
    x \equiv e^{- \varepsilon \mathcal{B}} = e^{- \varepsilon \beta}  e^{- i \varepsilon \tau} ,
    \label{x_def}
    \end{equation}
\noindent
which allows us to rewrite the partition function as a polynomial:
\begin{equation}
   Z(\mathcal{B},V,N) =   e^{- \mathcal{B} E_0} \sum_{k=0}^{n} \Omega_k x^{k}  =  e^{- \mathcal{B} E_0} \prod_{k=1}^{n} \left( x - x_{k}  \right),
   \label{polynomial}
\end{equation}
\noindent
  where $\Omega_k's$ are identified as the coefficients of the polynomial and $x_k$ represents the $k$-th zero. It is worth mentioning that a multiplicative constant in the polynomial does not alter its roots. Consequently, instead of dealing directly with the number of states, sometimes it can be preferable to work with the density of states (DOS), defined as $g(E) = \Omega(E) / \sum_E \Omega(E)$. 
  
According to the fundamental theorem of algebra, an $n$-th-order polynomial has exactly $n$ zeros, including multiplicities. Besides that, the roots of the polynomial occur in complex conjugated pairs ($x_{k_{\pm}} = e^{-\varepsilon \beta_k} e^{\pm i \varepsilon \tau_k}$). Since all coefficients in the polynomial are real and positive, any real zeros must be negative, at least for a finite-order polynomial. If $Z$ possesses real positive roots, the corresponding $F$ becomes singular at those points, indicating the presence of phase transitions in the system. Implying that a real positive zero is only possible at the thermodynamic limit. 
  
    All thermodynamic functions can be derived from the zeros, for instance, the specific heat at constant volume,
\begin{eqnarray} \nonumber
    c_V &=&  \frac{k_B\beta^2}{N} \left( \frac{\partial^2 \ln{Z}}{\partial \beta^2} \right)\\
    &=&  \frac{k_B x(\ln{|x|})^2}{N} \sum_{k=1}^{n} \left( \frac{-x_k}{(x-x_k)^2}\right).
    \label{cv_zeros}
\end{eqnarray}
\noindent
In this work, $c_V$ is measured in units of $k_B$. It is observed that a singular behavior of the specific heat may emerge for $x=x_{k}$ and $\tau_{k}  \ll 1$. While true phase transitions are not possible in finite systems, it is expected that a particular zero, known as the dominant or leading zero, approaches the real positive axis, indicating a pseudo-phase transition.

%
%
\subsection{Classification of the Order of the Phase Transition}
\label{Borrmann_section}

\begin{figure}[hbt!]
   \centering
     \includegraphics[width=0.4\textwidth,keepaspectratio=true,clip]{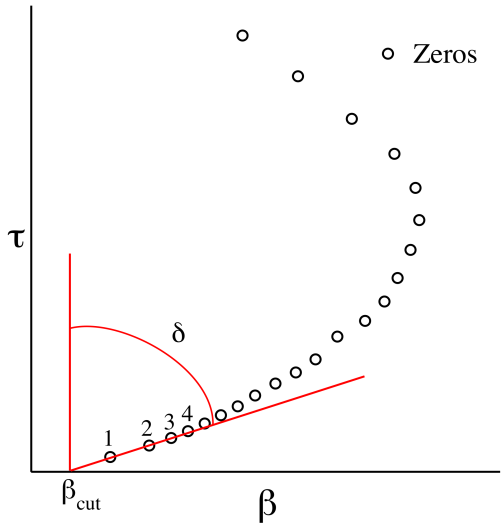}
     \caption{(Color online) Reproduction of the scheme of the DOZ toward the real axis from P.~Borrmann et al~\cite{Borrmann}
       \label{zerosFSS}}
\end{figure}

    P.~Borrmann et al~\cite{Borrmann} proposed a discretized version of the phase transition classification scheme of S.~Grossmann and W.~Rosenhauer~\cite{Grossmann1967,Grossmann1969}. In this section, we provide a brief outline of their main results. They considered the zeros that are close to the real axis to lie approximately on a straight line  making an angle $\delta = \arctan{(\gamma)}$ with the imaginary axis, where
\begin{equation}
    \gamma = \frac{\beta_2 - \beta_1}{\tau_2 - \tau_1},
    \label{gamma_eq}
\end{equation}
\noindent
as show in Fig.~\ref{zerosFSS}. It is worth mentioning that the indexes start from $1$ and increase with $\tau$, the zero labeled $1$ is the leading zero. The crossing point of the line with the real axis is $\beta_{cut} = \beta_1 - \gamma \tau_1$.
\noindent
\noindent
    A discrete density of zeros, $\phi(\tau_k)$, is defined as the average of the distances between the first near zeros as
\begin{equation}
    \phi(\tau_k) = \frac{1}{2}\left( \frac{1}{\|\mathcal{B}_{k} - \mathcal{B}_{k-1}\|} + \frac{1}{\|\mathcal{B}_{k+1} - \mathcal{B}_{k}\|}  \right),
    \label{phi_eq}
\end{equation}
\noindent
    with $k=2,3,4\cdots$. Since zeros with small imaginary parts contribute more to the specific heat at the transition (or any other thermodynamic functions that is singular at this point) they supposed that $\phi$ can be approximated by a  simple power law, i.e. $\phi(\tau) \sim \tau^{\alpha}$. An estimate of the exponent $\alpha$ can be done using two zeros as
\begin{equation}
\alpha = \frac{\ln{\phi(\tau_3)} - \ln{\phi(\tau_2)}}{\ln{\tau_3} - \ln{\tau_2} }.
\label{alpha_eq}
\end{equation}
\noindent
A first-order phase transition is defined by $\alpha = 0$ and $\gamma = 0$, i.e. a vertical line of evenly spaced zeros.
%
%
\section{Results}
\subsection{Fisher's zeros for a first-order phase transition}
  \label{firstOrderzeros}
In this section, we present an alternative demonstration that for a pseudo-first-order transition, the zeros of the partition function exhibit a vertical line pattern in the complex inverse temperature plane. We divide the domain of the partition function, given by eq.~(\ref{Z_total}), into three parts: $Z(\mathcal{B},V,N) = Z_{<} + Z' + Z_{>}$. The first part, $Z_{<}$, includes energies $E < E'$, $Z'$ covers the energy range of the non-stable region [$E'$, $E''$] (as discussed in section~\ref{microcanonical}), and $Z{>}$ accounts for energies $E > E''$.

It can be claimed that $Z'(\mathcal{B} =\mathcal{B}_j)  \approx 0$ since approaches that truncate the energy range, such as the zeros of the density of states~\cite{CMRocha1, Mota, Ronaldo_zeros}, can capture indications of phase transitions. Thus, we have:
\noindent
\begin{equation}
Z'(\mathcal{B}_j,V,N) =   \sum_{E=E'}^{E''} \Omega(E)e^{-\mathcal{B}_j E} \approx 0.
\label{Z2}
\end{equation}
\noindent

To deal with the convexity of the entropy, we expand $S$ in a Taylor series around the midpoint $E_{in} = (E' + E'')/2$ and collect terms up to the third order:
\begin{eqnarray} \nonumber
    S(E)  &\approx& S_{in} + \bar{\beta}_{in}(E  - E_{in}) + \frac{\bar{\gamma}_{in}}{2}(E -E_{in})^2 \\
          & +&  \frac{\bar{\delta}_{in}}{6}(E -E_{in})^3,
\label{S_taylor}
\end{eqnarray}
\noindent
    where  $S_{in} = S(E_{in})$ and $\bar{\beta}_{in} = \bar{\beta}(E_{in})$, $ \bar{\gamma}_{in} =  \bar{\gamma}(E_{in})$, and $ \bar{\delta}_{in} =  \bar{\delta}(E_{in})$ are the derivatives of $S$ as defined in section~\ref{microcanonical}.
In the considered energy range, $E  = E' + k\varepsilon$. Defining $\Delta E =   E''- E'$ so that $ E'=   E_{in} - \Delta E/2$, we can write:
\begin{equation}
    S(E) \approx  S' + \bar{\beta'} \varepsilon k + \frac{ \bar{\gamma'} }{2}  \varepsilon^2 k^2 + \frac{\bar{\delta}_{in}}{6} \varepsilon^3k^3,
\label{Sk}
\end{equation}
\noindent
    where,
\begin{equation}
    S' =  S_{in} - \frac{\bar{\beta}_{in}}{2}\Delta E + \frac{\bar{\gamma}_{in}}{8}\Delta E^2 - \frac{\bar{\delta}_{in}}{48}\Delta E^3,
\end{equation}
\noindent
\begin{equation}
    \bar{\beta'} = \bar{\beta}_{in} - \frac{\bar{\gamma}_{in}}{2}\Delta E + \frac{\bar{\delta}_{in}}{8}\Delta E^2 =- \frac{\partial S'}{\partial E'},
\label{temp_correction}
\end{equation}
\noindent
    and
\begin{equation}
    \bar{\gamma'} =  \bar{\gamma}_{in} - \frac{\bar{\delta}_{in}}{2}\Delta E = -\frac{\partial \bar{\beta'} }{\partial E'} = \frac{\partial^2 S'}{\partial E'^2} .
\end{equation}
\noindent
    Inserting eq.~(\ref{Sk}) into eq.~(\ref{PlanckBoltzmann}) and solving for $\Omega(E)$, eq.~(\ref{Z2}) can be rewritten as:
\[
Z'_n(\mathcal{B}_{j}) \approx e^{-\mathcal{B}_{j}F'}\sum_{k=0}^{n'} x^ky^{k^2}z^{k^3},
\]
\noindent
    where  $n'$ is the number of energy levels in the energy range of the non-stable region, $F'= E' - S'/(k_B\mathcal{B}_{j})$,
\begin{eqnarray} \nonumber
x  &=& \exp{\left[- \left(\mathcal{B}_{j} - \frac{ \bar{\beta'}}{k_B} \right)\varepsilon \right]} \\
  &=& \exp{\left[- \left(\beta_{j} - \frac{\bar{\beta'}}{k_B} \right)\varepsilon \right]}   \exp{\Big[ -i \tau_{j} \varepsilon \Big]} ,
\label{variableX}
\end{eqnarray}
\noindent
\[
y = \exp{\left(\frac{\bar{\gamma'}  }{2k_B} \varepsilon^2 \right)},
\]
and
\[
z = \exp{\left( \frac{\bar{\delta}_{in} }{6k_B} \varepsilon^3\right)}.
\]
\noindent
    Usually, $\varepsilon$, $\bar{\gamma}_{in}$ and $\bar{\delta}_{in}$ are small quantities, so $y\approx z \approx 1$ giving:
\begin{equation}
    Z' \approx e^{-\mathcal{B}_{j}F'}\sum_{k=0}^{n'} x^k =e^{-\mathcal{B}_{j}F'} \frac{1 - x^{n'+1}}{1 - x}.
\label{Z2map}
\end{equation}
\noindent
    By collecting terms up to first order, i.e. considering a linear behavior of the entropy, it leads to the same relation for $Z'$. Hence, one can say that the double-touching tangent line construction is a good approach even for finite systems. By inspecting eqs.~(\ref{Z2map}) and (\ref{variableX}), we get $Z'= 0$ if
\begin{equation}
    \beta_{j} = \frac{\bar{\beta'}}{k_B},
\label{betaCanonico_micro}
\end{equation}
\noindent
    and
 \begin{equation}
   \tau_{j} = \frac{2\pi j}{ \varepsilon (n'+1)} =  \frac{2\pi }{ \Delta E}\ j
   \label{tau_j}
 \end{equation}
 \noindent
     where $j = 1, 2, \cdots, n'$. It is worth mentioning that $j \ne 0, (n'+1)$, since the denominator in the last term of eq.~(\ref{Z2map}) requires that $x \ne 1$, hence $\mathcal{B}_{j}$ can not be a positive real number. Furthermore, any other $j$ will lead to multiplicities and can be neglected. Since $\bar{\beta'}$ is a constant, given by eq.~(\ref{temp_correction}), plotting the ordered pairs ($\beta_{j}$, $ \tau_{j}$) leads to a vertical line of evenly spaced points as claimed before. Besides that, by inserting eq.~(\ref{temp_correction}) into eq.~(\ref{betaCanonico_micro}), we obtain:
\begin{equation}
  k_B\beta_{j} = \bar{\beta}_{in}  - \frac{\bar{\gamma}_{in}}{2}\Delta E  + \frac{\bar{\delta}_{in}}{8}\Delta E^2.
  \label{betaCorrection}
\end{equation}
%
%
\subsection{Zeros Map for the Lennard-Jones Cluster}
  \label{LJclusterZeros}

\begin{figure}[hbt!]
\centering
      \includegraphics[width=0.7\textwidth,keepaspectratio=true,clip]{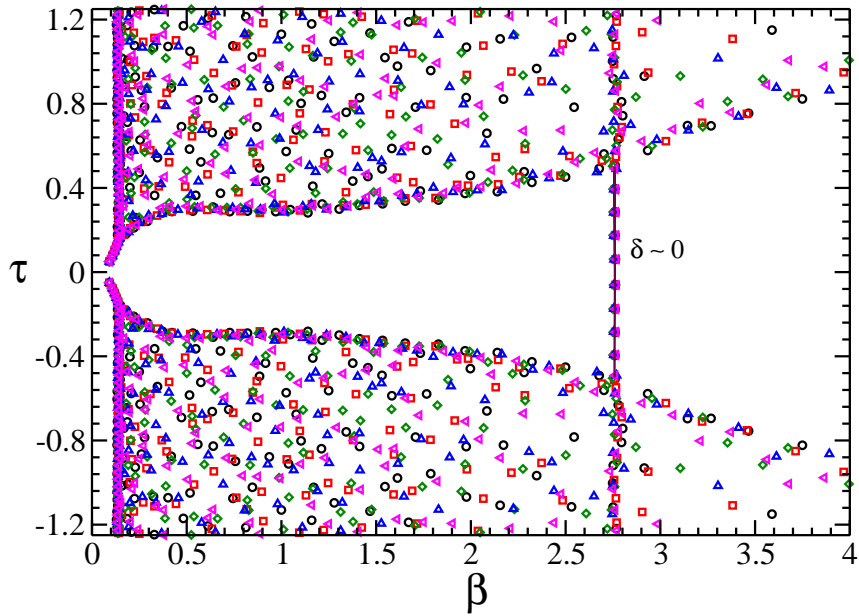}
    \caption{(Color online)  The Fisher zeros distribution map for the $147$ particles Lennard-Jones cluster. Each symbol indicates the results of an independent simulation. $\delta$ is the angle between the fitted line of the dominant zeros with the vertical axis. See the inset of Fig.~\ref{scallingLJ} for a zoom in this region.
        \label{LJzerosBetaMap}}
\end{figure}

In this section, we illustrate the discussion above by using the example of the Lennard-Jones (LJ) cluster with $N=147$ particles, which is considered a prototype of a pseudo-first-order phase transition. The LJ cluster consists of particles interacting through the pairwise LJ potential:
\begin{equation}
    U_{LJ}(r_{ij}) = 4\epsilon \left[ \left(\frac{\sigma}{r_{ij}}\right)^{12} -  \left(\frac{\sigma}{r_{ij}}\right)^6 \right],
\end{equation}
where $r_{ij} = |\mathbf{r}_j - \mathbf{r}_i|$ is the distance between particles identified by the indices $i$ and $j$, and $\mathbf{r}_i$ and $\mathbf{r}_j$ denote the respective positions of these particles. The reduced parameters are set such that the minimum of the potential is at $r_{ij}=r_0=1$, and energy is measured in units of $\epsilon$ ($\sigma = 2^{-1/6}$ and $\epsilon = 1$). The particles are confined to a sphere of radius $r_c = 4\sigma$ to reproduce the transition temperature ($T_{tr}\approx 0.36$) presented by P.A.~Frantsuzov and V.A.~Mandelshtam\cite{Frantsuzov}. The results presented in this section are averages of five independent simulations, and errors are given by the standard deviation, except for Fig.~\ref{LJzerosBetaMap}, where the zeros map of each individual simulation is shown. The specific details of the simulations can be found in Appendix~\ref{rewl_details}.

Fig.~\ref{LJentropy} shows the specific entropy, $s = S/N$, as a function of the energy density, $e = E/N$, for the $147$-LJ cluster. One can observe the convex intruder inside the dotted green rectangle, which is zoomed in the inset. The blue dashed line is the double-touching tangent line construction, which leads to a slope $\bar{\beta}_{tan} = 2.751(9)$, and the energy density range of the unstable region is $[e' = -5.2286(9), e'' = -4.861(1)]$. The specific latent heat then is calculated to be $q_L = 2.78(1)$.
\noindent
\begin{figure}[hbt!]
    \centering
      \includegraphics[width=0.7\textwidth,keepaspectratio=true,clip]{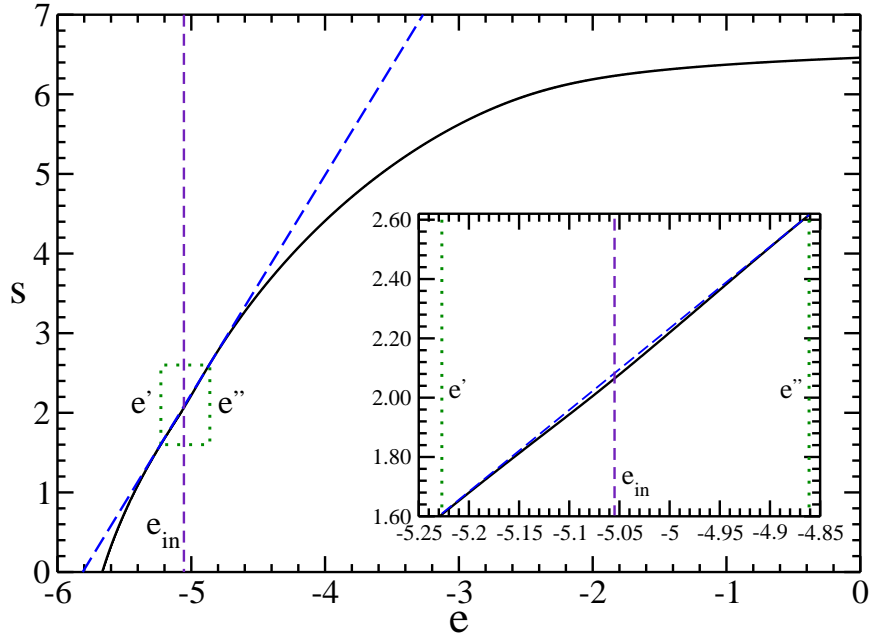}
    \caption{(Color online) Estimation of the specific entropy for the $147$ particles Lennard-Jones Cluster. The error bars are in the same order as the line width. The dotted green rectangle demarcates the unstable region. The inset is a zoom in this region where the convex intruder can be perceived. The dashed blue line is the double-touching tangent line construction. The small dashed purple vertical line marks $e_{in} = (e'' + e')/2$.
        \label{LJentropy}}
\end{figure}
\noindent

\begin{figure}[hbt!]
\centering
      \includegraphics[width=0.7\textwidth,keepaspectratio=true,clip]{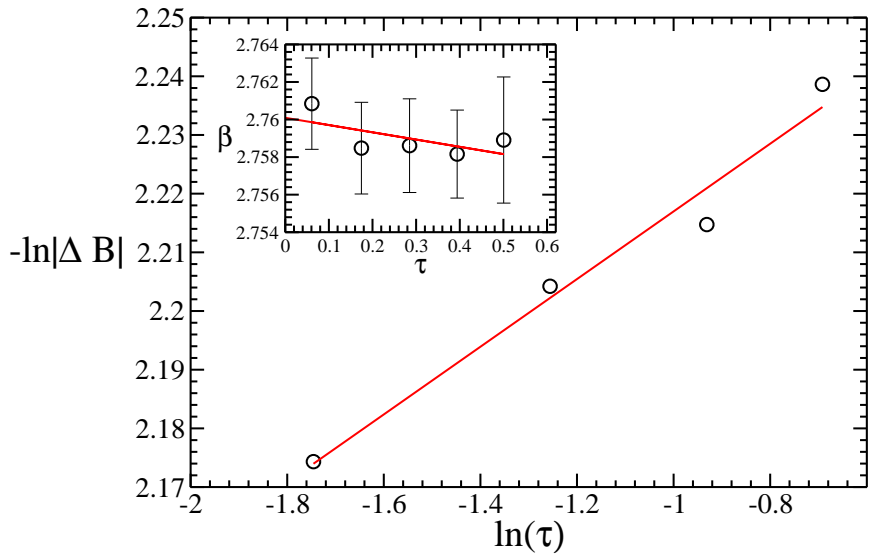}
    \caption{(Color online) $\log \times \log$  graph of the inverse of the absolute value of the difference between the complex inverse temperature of adjacent dominant zeros versus the complex part of the inverse temperature, i.e. $-\ln{\|\mathcal{B}_k - \mathcal{B}_{k-1}\|} \times \ln{(\tau_k)}$, for $k=2,3,4$, and $5$. In the inset we show the real part versus the imaginary part of the dominant zeros.
        \label{scallingLJ}}
\end{figure}

Fig.~\ref{LJzerosBetaMap}  displays the Fisher zeros distribution map for the $147$-LJ cluster. Each symbol indicates the result of an independent simulation. The leading zero is found at ($\beta_1 = 2.761(2)$, $\tau_1 = 0.0609(6)$), our maps also show that although the zeros are sensitive to statistical fluctuations, the zeros in the transition region are quite stable~\cite{Schnabel_zeros}. In Fig.~\ref{scallingLJ} we show an adaptation of the scaling analysis proposed by Borrmann et al, discussed in section~\ref{Borrmann_section}. We propose a linear fit in $\ln{(1/\|\mathcal{B}_k - \mathcal{B}_{k-1}\|)} \times \ln{(\tau_k)}$, for $k=2,3,4,$ and $5$. We found the coeficient $\alpha = 0.058(7)$, which is coherent to the approach proposed by eq.~(\ref{alpha_eq}),  $\alpha = 0.041(5)$. In the inset of this figure we show the linear fit of the dominant zeros where we found the slope $\gamma = -0.004(3)$ which leads to an angle  $\delta = 0.2(2)^{\circ}$, and the crossing point $\beta_{cut} = 2.7601(9)$. Those parameters are consistent with the first-order phase transition. Besides that, they are also consistent with the approach values proposed by Borrmann et all, $\gamma = -0.021(1)$, and $\beta_{cut} = 2.762(2)$. The average of the distances between the dominant zeros is $0.110(2)$. From eq.~(\ref{tau_j}) one can see that this distance is $\Delta \tau = 2\pi/[N(e''-e')] = 0.1162(4)$, corroborating for the validity of the demonstration.
\noindent
\noindent

\noindent
\noindent

\begin{figure}[hbt!]
\centering
      \includegraphics[width=0.7\textwidth,keepaspectratio=true,clip]{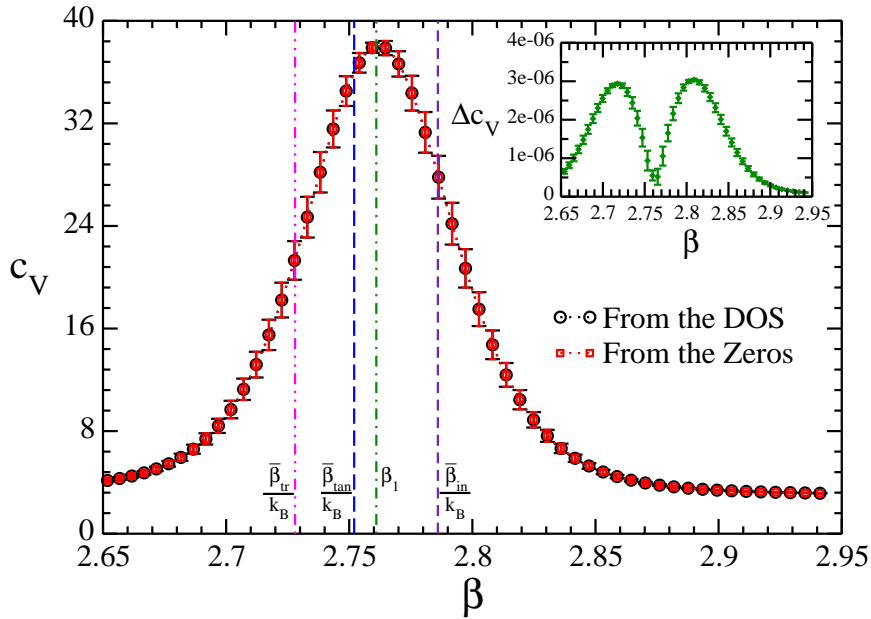}
      \caption{(Color online) Specific heat at constant volume for the $147$ Lennard-Jones Cluster ($V = 4^4 \pi \sigma^3/3$). The black circles stands for $c_V$ evaluated via the DOS, eq.~(\ref{cvDOS}). The red square stands for $c_V$ calculated via the Fisher zeros, eq.~(\ref{cv_zeros}). The inset shows the relative difference between the two values, see eq.~(\ref{relativeDiff}). The dotted-dashed green line indicates $\beta_{1}$  from the zeros maps, the dashed blue line indicates $\bar{\beta}_{tan}$ from the tangent line of double-touching tangent line construction, and the small dashed purple line indicates $\bar{\beta}_{in}$ and the double-dotted-dashed magenta line indicates $\bar{\beta}_{tr}$, which will be discussed later.
        \label{cv_LJ}}
\end{figure}

The reliability of the zeros maps and their relationship with other quantities is further discussed. We have chosen the MPSolve~\cite{mpsolve1,mpsolve2} routine as the zeros finder for this study. Besides the roots of polynomials, this routine's output can also return error bars. In this examination, the error bars are the order of $10^{-12}$. Upheld by obtaining $\sum_ i \tau_ i \approx 0$, since the zeros come in complex conjugated pairs, we can endorse the precision of the routine in this case. To prove accuracy, one can calculate a given thermodynamic function by the Fisher's zeros and compare it with one obtained via DOS. As a check, we compare the specific heat at constant volume obtained by eq.~(\ref{cv_zeros}) and by the standard canonical average:
\begin{equation}
    c_V = \frac{k_B\beta^2}{N}\left( \left\langle E^2 \right\rangle - \left\langle E \right\rangle^2  \right),
\label{cvDOS}
\end{equation}
    where
\begin{equation}
    \left\langle E^k \right\rangle =  \sum_E E^k P(E,\beta),
\label{canonicalAverage}
\end{equation}
    and
\begin{equation}
    P(E,\beta)= \frac{ g(E)e^{-\beta E}}{Z} ,
\label{P_Boltzmann}
\end{equation}
    is the Boltzmann probability density. We then define the relative difference,
\begin{equation}
\Delta c_V =  \left\| 1 - \frac{c_V(z)}{c_V(g)} \right\|,
\label{relativeDiff}
\end{equation}
where $c_V(g)$ is obtained from the DOS and $c_V(z)$ is obtained from the zeros, as comparative metric. This inspection is shown in Fig.~\ref{cv_LJ}, where we can state that the numerical imprecision provided by the zeros finder is negligible in this case. Thus, we have high confidence in the legitimacy of the zeros map. In addition, one can recognize that the $\beta_1$, indicated by the dotted-dashed green line, is close to the temperature of the peak position of the $c_V$.

\noindent
\noindent

\begin{figure}[hbt!]
\centering
      \includegraphics[width=0.7\textwidth,keepaspectratio=true,clip]{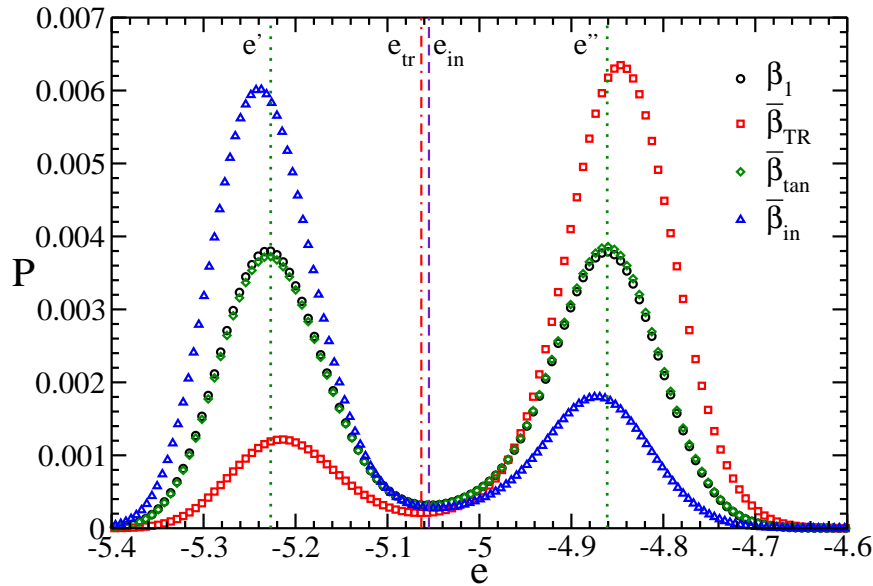}
       \caption{(Color online)  The Boltzmann probability density of the $147$ Lennard-Jones Cluster. The dotted-double-dashed red vertical line marks the microcanonical transition point. The unstable region is demarcated by the dotted green line.
        \label{boltzmann_graph}}
\end{figure}

    Due to the coexistence of phases, the Boltzmann probability density presents two peaks in a first-order transition, each related to a phase. At the transition temperature, one expected that those peaks have the same height. Since one can rewrite eq.~(\ref{P_Boltzmann}) as $P(E,\beta) = \exp{(-\beta F)}/Z$, this analysis is similar to the minimum condition of the Helmholtz free energy. In Fig.~\ref{boltzmann_graph} we show the Boltzmann probability density for four temperatures:  $\beta_{1}$ and $\bar{\beta}_{tan}$, already estabilished, and $\bar{\beta}_{in}$ and $\bar{\beta}_{tr}$, discussed in the next paragraph. One can see that the Fisher zeros analysis is coherent with the equal probability condition, and the double-touching tangent construction slightly deviates from it.
\noindent
\noindent

\begin{figure}[hbt!]
\centering
      \includegraphics[width=0.7\textwidth,keepaspectratio=true,clip]{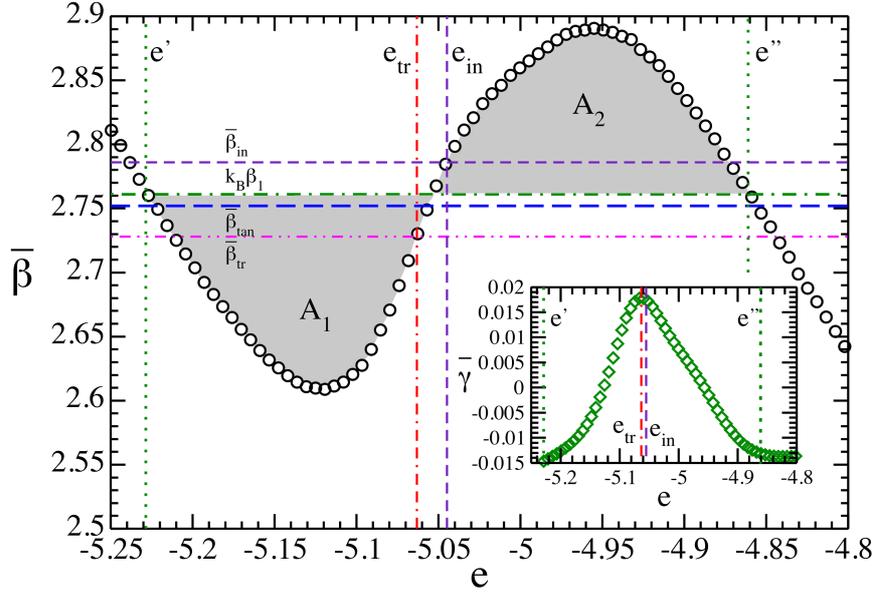}
          \caption{(Color online)  The microcanonical inverse temperature in the unstable region. The dashed blue line indicates $\bar{\beta}_{tan}$ from the double-touching tangent line construction, the dotted-dashed green line indicates $k_B\beta_{1}$  from the zeros maps, the small dashed purple line indicates $\bar{\beta}_{in} = \bar{\beta}(e_{in})$, and the double-dotted-dashed magenta line indicates $\bar{\beta}_{tr} = \bar{\beta}(e_{tr})$ from the microcanonical analysis. The error bars are the same size as the symbols. The hued areas $\mathrm A_ 1$ and $\mathrm A_ 2$ are consistent with the  Maxwell’s equal area construction. The inset shows $\bar{\gamma}$. The dotted-double-dashed red line marks the peak position of $\bar{\gamma}$, i.e. the microcanonical transition point.
        \label{beta_micro}}
\end{figure}

    Finally, we show the microcanonical analysis of least-sensitive inflection points for the $147$-LJ cluster. The change in the curvature of $S(E)$ causes an inflection point, called the inflection point of least sensitivity if the derivative changes least on variation in energy and provides a signal of the transition at this energy, $E_{tr}$~\cite{QiBachmann}. Let $\bar{\beta}_{tr} = \bar{\beta}(E_{tr})$, $ \bar{\gamma}_{tr} =  \bar{\gamma}(E_{tr})$, and $ \bar{\delta}_{tr} =  \bar{\delta}(E_{tr})$ the higher order derivatives of $S$ evaluated in $E_{tr}$. According to the least-sensitive inflection point microcanonical analysis, for a pseudo-first-order transition, $\bar{\gamma}_{tr}$ is a maximum positive value. In Fig.~\ref{beta_micro} we show the microcanonical inverse temperature, $\bar\beta$, just for the unstable region, i.e. the derivative of the entropy shown in the inset of Fig.~\ref{LJentropy}. In conformity, the dashed blue line is the derivative of the double-touching tangent line construction. For comparison purposes, we show $k_B\beta_1$ in the dotted-dashed green line. We also show $\bar{\beta}_{in}= \bar{\beta}(e_{in})$ in the small dashed purple line. $\bar{\beta}_{in} >k_B\beta_1$ as predicted by eq.~(\ref{betaCorrection}).  $k_B\beta_1$ line is in accordance with the Maxwell’s equal area construction, since $A_1 \approx A_2$.  In the inset, we show $\bar{\gamma}$, measured in units of $k_B/\epsilon^3$, where the peak position defines the microcanonical transition point, $e_{tr}$. The double-dotted-dashed magenta line indicates the microcanonical transition temperature, i.e.  $\bar\beta_{tr} =\bar{\beta}(e_{tr})$.

\noindent
\noindent

    It is worth mentioning that, although the Fisher zeros analysis corroborates with the Maxwell’s equal area construction, equal probability condition, and provides a transition temperature close to the temperature of the peak position of the specific heat. It is well known that, for finite systems, different quantities provides different transitions temperatures~\cite{Schnabel_zeros}, converging to the transition value as the thermodynamic limit is approached. Thus, this specific study is inconclusive about the accuracy of distinct methods, a statement in this regard requires extensive work, and this is not the purpose of this manuscript.

%
%
\section{Conclusion}
  \label{Final_Remarks}
In this work, a mathematical argument to connect the nonstable region of the entropy to the pattern of the Fisher's zeros map was presented. 
The first term of Taylor's series approach of the entropy leads to the vertically lined equally spaced zeros of the partition function on the complex inverse temperature plane for the first-order phase transition. This behavior was illustrated via the Lennard-Jones cluster.  For this specific model, the transition temperature defined by the leading zero corroborates with the peak position of the specific heat, also with the Maxwell’s equal area construction, and with the equal probability condition of phases at the first-order transition.

%
%
\section*{Acknowledgments}
We would like to acknowledge helpful conversations with Dr. Michael Bachmann. This work received public financial support from Fundação de Amparo à Pesquisa do Estado de Minas Gerais (FAPEMIG), Brazil, under the grant RED-00458-16.

%
%
\section*{Declarations}
 The authors have no competing interests to declare that are relevant to the content of this article.
%
%

\appendix
\section{Details of the Simulations}
\label{rewl_details}

In this appendix, we present a detailed description of the Monte Carlo simulation method used for studying the Lennard-Jones cluster. 
The Monte Carlo method is a class of statistical algorithms that sample a limited but representative number of states to infer some properties of the system under study. Those states can be chosen in a Markov chain, i.e. the probability to sample each state depends only on the previous state. Mathematically, this condition can be stated by the detailed balance,
\begin{equation}
P_{i}W_{i\to j}=P_{j}W_{j \to i},
\end{equation}
where $W_{i\to j}$ is the transition probability from state $i$ to state $j$, and $P_i$ is the equilibrium probability of being in state $i$~\cite{landau_book}.
The Metropolis prescription to satisfy this condition is
\begin{equation}
\text{W}_{i\to j} =\text{min}\left\{1, \frac{P_j}{P_i}\right\}.
\end{equation}

We want a Monte Carlo scheme to estimate the entropy, it can be done by one of the flat histogram methods, here we choose the Wang-Landau Sampling~\cite{Fugao}. To understand this method, let us look at the Boltzmann distribution for $\beta = 0$. In this situation, eq.~(\ref{P_Boltzmann}) can be written as $P(E) = \Omega(E)/Z$. So, the probability of randomly tossing a state with energy $E_i$ is proportional to $\Omega(E_i)$. If we accept the tossed state to our sampled set with probability $P_i = 1/\Omega(E_i)$, all energies will be equally sampled. Of course, we are unaware of $\Omega(E)$, but we can use this equally sampled energies fact to estimate it as follows: We create a histogram to count how many states with a given energy were sampled, $h(E)$. Since $\Omega(E)$ can assume very large numbers, let us work with the entropy. So we first guess an initial value to $S(E)/k_B$, for instance, $\ln(\Omega(E)) = 1$, and define an initial state, $i$. Hereinafter, we randomly guess a new state, $j$, and compare the states $i$ and $j$ by the Metropolis prescription. Considering the proposed probability, it can be written as
\begin{equation}
\text{W}_{i\to j} =\text{min}\left\{1, \frac{\Omega(E_i)}{\Omega(E_j)}\right\}.
\end{equation}
If the trial state is accepted we set $j$ as the current one, $i\leftarrow j$. At every trial move $\Omega(E_i)$ is updated by a multiplicative factor $f$, i.e., $\ln{\Omega(E_i)} \leftarrow \ln{\Omega(E_i)} + \ln{(f)}$. Simultaneously,  the histogram is also updated, $h(E_i) \leftarrow h(E_i) + 1$. When $h(E)$ is flat we can say that we approach $\ln{\Omega(E)}$ with precision equal to $\ln(f)$.  We considered the energy ranging from $ 0.95 E_{\text{min}}$ to $E_{\text{max}}=0$. Where $E_{\text{min}}$ is the ground state given by J.A. Northby~\cite{Northbya}. We also consider one trial move the attempt to change the position of a single particle. The new position is chosen inside a small sphere of radius $r_t$ centered in the original position of the particle. The value of $r_t$ is chosen so that the acceptance ratio is close to $60\%$.
To quickly sample the entire configuration space, a large initial value for $f = f_0$ is required, the original recommendation states that $\ln{(f_0)} = 1$. To improve the precision $f$ must be decreased and the scheme repeated.

\begin{figure}[hbt!]
\centering
      \includegraphics[width=0.7\textwidth,keepaspectratio=true,clip]{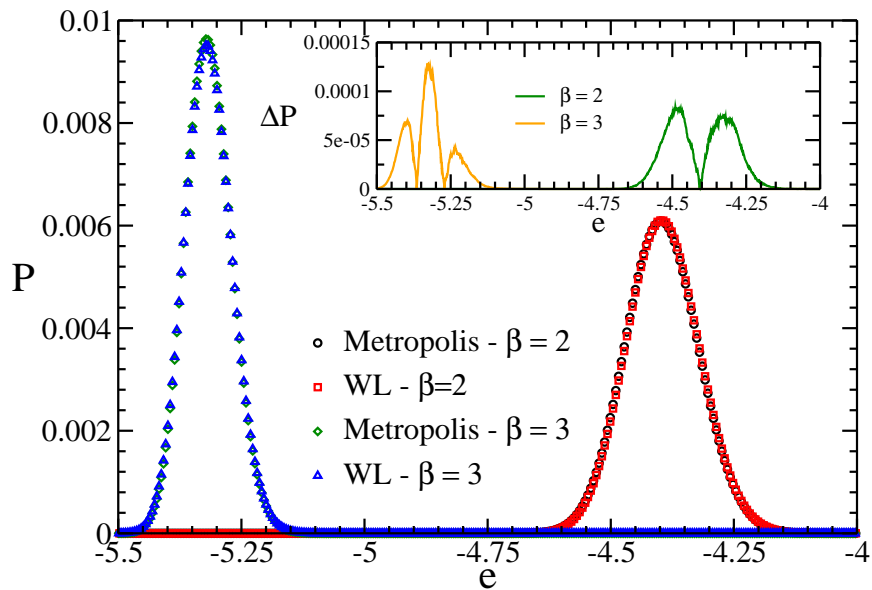}
       \caption{(Color online)  Comparison of the Boltzmann probability density of the $147$ Lennard-Jones Cluster for $\beta = 2.0$ and $\beta = 3.0$ obtained by the Replica-Exchange-Wang-Landau method and by the Metropolis algorithm.
        \label{boltzmann_check}}
\end{figure}

The histogram flatness is tested after $10^6$ Monte Carlo sweeps (MCS). One MCS is counted after a sequential attempt to change all particles of the system once. If the histogram is flat, it is reset, $h(E) = 0$ and $f$ is decreased. The histogram is considered flat when the ratio of its lowest value by the mean value is greater than $p$, in this work $p=0.70$. At first, any function can be used to decrease $f$, we also used the original suggestion, i.e. $\ln{(f_{i+1})} = \ln{(f_i)}/2$. The scheme is repeated until the desired precision is reached, in this work we cease the process when $\ln{(f)} = \varepsilon = 10^{-9}$. Regrettably, the capacity to diminish the inaccuracy of $\Omega(E)$ asymptotically halts as the modification factor $f$ decreases. This phenomenon is recognized as the saturation of the error between the calculated and the exact $\Omega(E)$, a concept first elucidated in Ref.~\cite{qiliang}. Various improvements to the WL-method have been proposed, for instance, the $1/t$-Wang-Landau approach~\cite{belardinelli1,caparica,bb} and the optimal modification factor~\cite{chenggang}. Given the analogous behavior observed in the convergence patterns of simple sampling Monte Carlo and $1/t$-Wang-Landau~\cite{belardinelli2}, we compare our results with those obtained by the regular Metropolis algorithm~\cite{Metropolis}, as illustrated in Fig.~\ref{boltzmann_check}. The discussion of this comparison is presented in the final paragraph of this appendix.

Moreover, the standard WL method is very time-consuming, so we opted for a parallelization procedure, called Replica Exchange Wang-Landau (REWL) method \cite{REWL}. The idea is to divide the energy range into several smaller pieces, called windows. In this work, we divided the energy range in $N_{W} = 10$ equal size windows of $10^4$ energy bins. One or more WL samples, called walkers, are performed in parallel at each window. Here we consider $N_{R} = 4$ walkers in each window. In addition, an attempt to exchange configurations of walkers between adjacent windows is proposed after $10^3$ MCS. An exchange between conformations $X$ and $Y$, respectively located at neighboring windows $i$ and $j$, is proposed with the probability
 \begin{equation}
     \text{P}_{\text{acc}} =\text{min}\left\{\frac{\Omega_i(E[ X] )}{\Omega_i(E[ Y] )}\frac{\Omega_j(E[ Y] )}{\Omega_j(E[ X] )},1\right\} .
 \end{equation}
 This exchange allows the walkers to efficiently sample different parts of the configuration space, this procedure is as crucial as dividing the windows to improve the simulation time. The acceptance ratio of the replica exchange is tied to the overlap between the windows, in this work we set an overlap of $75\%$.  When the final precision is reached, the pieces are combined to form the entire entropy. We concatenate the pieces at the point of the smallest difference of the inverse temperature between the adjacent windows. There are $N_{R}^{N_{W}}$ possible combinations of concatenations of the pieces of the entropy, we randomly chose $10^3$ of them and the final result is the average value of those combinations via Jackknife resampling. The average procedure presented here holds significant importance as it effectively mitigates the potential introduction of spurious non-analytical points in the entropy, arising from the interconnection of distinct entropy fragments. These artifacts, if not properly handled, might erroneously suggest the presence of non-existent phase transitions. Since each pair of walkers is connected at a different point, the implementation of the Jackknife resampling technique induces a larger weight in the continuous region of the pieces of the entropy, which ultimately improves the accuracy of the mean value.

On the question of the convergence issue of the Wang-Landau method we check the Boltzmann distribution obtained by the REWL method with the one obtained by the regular Metropolis Algorithm~\cite{Metropolis}, see Fig.~\ref{boltzmann_check}. We calculate the $P(E,\beta)$ for two temperatures, one above the transition temperature ($\beta = 2$) and another below ($\beta = 3$). Those temperatures are far away from the transition to avoid the Metropolis algorithm being stuck in a meta-stable state~\cite{MUCA}. For the Metropolis Algorithm, we excluded the first $10^5$ MCS for thermalization purposes and, after that, performed $10^7$ MCS to obtain $P(E)$. The result presented here is also an average of 5 independent simulations. Besides that, the trial move is similar to that used for the WL method. The relative differences between the two methods are of the order of the error bars, see the inset in Fig.~\ref{boltzmann_check}, demonstrating the reliability of the REWL procedure.

\noindent
\noindent

\bibliographystyle{unsrt}  
\bibliography{pattern_zeros_arxiv}

\end{document}